\documentclass{ptephy}
\usepackage{amsmath,amssymb}
\usepackage{bm}
\usepackage[dvipdfmx,implicit=false,colorlinks=true,linkcolor=blue]{hyperref}

\newcommand{\eq}{\begin{equation}}
\newcommand{\eqend}{\end{equation}}
\newcommand{\ovl}{\overline}
\newcommand{\A}{\alpha}

\begin{document}

\title{Supergravity on the noncommutative geometry}

\author{\name{\fname{Masafumi} \surname{Shimojo}}{1,\ast}, \name{\fname{Satoshi} \surname{Ishihara}}{2}, 
\name{\fname{Hironobu} \surname{Kataoka}}{2}, \name{\fname{Atsuko} \surname{Matsukawa}}{2}\\ and  \name{
\fname{Hikaru} \surname{Sato}}{2}
}

\address{
\affil{1}{Department of Electronics and Information Engineering, National Institute of Technology, Fukui College, 
Geshicho, Sabae, Fukui 916-8507, Japan}
\affil{2}{Department of Physics, Hyogo University of Education, Shimokume, Kato, Hyogo 673-1494, Japan}
\email{shimo0@ei.fukui-nct.ac.jp} \\
}

\begin{abstract}%
Two years ago, we found the supersymmetric counterpart of the spectral triple which specified noncommutative 
geometry. Based on the triple, we derived gauge vector supermultiplets, Higgs supermultiplets of the minimum 
supersymmetric standard model and its action. However, unlike the famous theories of Connes and his co-workers, 
the action does not couple to gravity. In this paper, we obtain the supersymmetric Dirac operator 
$\mathcal{D}_M^{(SG)}$  on the Riemann-Cartan curved space replacing derivatives which appear in that 
of the triple with the covariant derivatives of general coordinate transformation. We apply the supersymmetric 
version of the spectral action principle and investigate  
the heat kernel expansion on the square of the Dirac operator. As a result, we obtain a new supergravity 
action which does not include the Ricci curvature tensor. 
\end{abstract}
\subjectindex{B11, B16, B82}
\maketitle
\section{Introduction}
The standard model of high energy physics has some defects. It can not include gravity theory, can not 
solve hierarchy problem, has many free parameters including coupling constants of gauge groups in the 
theory which must be decided by experiments. More essentially, it cannot explain why the gauge group 
is $SU(3) \times SU(2) \times U(1)$.    
 Connes and his co-workers derived the standard model coupled to gravity on the basis of noncommutative  geometry(NCG)\cite{rf:Grcoupled,connes1,rf:JHigh,rf:NCGneutrino}. Their result is that if 
the space-time is a product of a 
continuous Riemannian manifold $M$ and a finite space $F$ of KO-dimension 6, gauge theories of the 
standard model are uniquely derived\cite{connes3,conceptual}. 
In the model, three coupling constants of $SU(3)$, $SU(2)$, $U(1)$ are unified by the same relation 
as that of $SU(5)$ grand unified theory(GUT). The Weinberg angle is also fixed at that of the GUT. 

On the other hand, the most powerful candidate of new physics to solve the hierarchy problem 
is supersymmetric theory\cite{susy}. One loop correction to squared Higgs mass $m_H^2$ from a 
Dirac spinor contains square of ultraviolet cut off, $\Lambda_{UV}^2$. If $\Lambda_{UV}$ is 
order of Plank scale, this correction is 30 orders of magnitude larger than the value of $m_H^2$. 
Introducing supersymmetry brings about one more loop correction from a boson which is the superpartner of the 
fermion. It has the same absolute value as that of the fermion loop but with opposite sign, so that the hierarchy problem 
is systematically removed.  

Unfortunately, it is difficult, perhaps impossible to extend the NCG itself to new one which produce 
supersymetric particle models. The framework of NCG is specified by so-called the spectral triple 
$(\mathcal{H}_0 ,\mathcal{A}_0,\mathcal{D}_0)$, where $\mathcal{H}_0$ is a Hilbert space which consists of 
spinorial wave functions of matter fields in the standard model, $\mathcal{A}_0$ and $\mathcal{D}_0$ are 
algebra and Dirac operator which act on $\mathcal{H}_0$. Let us consider to extend the spectral triple 
to supersymmetric counterpart, $(\mathcal{H},\mathcal{A},\mathcal{D})$, where $\mathcal{H}$
is the space which consists of not only spinorial wave functions but also of bosonic wave functions
of superpartners of the matter fields and elements of $\mathcal{A}$ and $\mathcal{D}$ are 
operators which act on $\mathcal{H}$. 
Supersymmetric theories are mostly formulated on the Minkowskian space-time, while the standard model constructed 
from the NCG is formulated on the space-time with Euclidean signature. So the space $\mathcal{H}$ will not be a
Hilbert space. In addition, since the extended Dirac operator $\mathcal{D}$ will include d'Alembertian 
which appears in the Klein-Gordon equation, $[\mathcal{D},a]$ will not be bounded for an arbitrary element $a\in \mathcal{A}$, 
so that $\mathcal{D}^{-1} $ can not play the role of infinitesimal length element $ds$ of a geometry. 
These facts do not obey axioms of NCG.  

Nevertheless, if not only supersymmetry but also NCG have important meaning in particle physics, 
these successful theories must coexist. Recently, we arrived at the minimum supersymmetric standard 
model(MSSM) based on NCG\cite{paper0,paper1,paper2}. We have found the supersymmetric counterpart 
of the process to construct the standard model action from the NCG one by one. 
At first, we obtained "the triple"  $(\mathcal{H},\mathcal{A},\mathcal{D})$ extended 
from the spectral triple and verified the supersymmetry of it. 
The space $\mathcal{H}$ is the product of 
the functional space $\mathcal{H}_M$ on the Minkowskian space-time manifold and the finite space  
$\mathcal{H}_F$ which is the space of labels denoting the matter particles. 
According to the constitution of $\mathcal{H}$, the algebra $\mathcal{A}$/the Dirac operator $\mathcal{D}$ 
consists of $\mathcal{A}_M$/$\mathcal{D}_M$ which acts on the manifold and $\mathcal{A}_F$/$\mathcal{D}_F$ 
which acts on the finite space, respectively. 
The above construction was performed in the Minkowskian signature in order to incorporate supersymmetry.
As mentioned earlier, the spectral triple corresponds to the NCG, but the triple does not define a new NCG. 
However, projecting $\mathcal{H}_M$ to the fermionic part and changing the signature 
from the Minkowskian one to the Euclidean one by the Wick rotation, we found that it reduced to the theory 
constructed on the original spectral triple. We derived internal fluctuation of the Dirac operator,  
$\widetilde{\mathcal{D}}$, which induced vector supermultiplets of gauge degrees of freedom and 
Higgs supermultiplets. 

Secondarily, we obtained the action of the NCG model in terms of the supersymmetric version of 
spectral action principle\cite{connes8} which was expressed by  
\begin{equation}
\sum_k(\Phi_k,{\rm i}\widetilde{\mathcal{D}} \Phi_k)+Tr_{L^2}f(P), \label{actionprinciple}
\end{equation}
where $\Phi_k$ denoted the wave functions which described the chiral or antichiral supermultiplet, 
$P=(i\tilde{\mathcal{D}})^2$ and $f(x)$ was an auxiliary smooth function on a 4D compact 
Riemannian manifold without boundary\cite{connes8}. We calculated the Seeley-Dewitt coefficients 
due to the  second term of (\ref{actionprinciple}), which gave the action of the Non-Abelian 
gauge fields and Higgs fields. 

However, since the triple is constructed on the Minkowskian space-time,i.e. 
flat space-time, it does not give the action of gravity. In this paper, 
We replace the derivative $i\partial_\mu$ which appears in $\mathcal{D}_M$ with 
the covariant derivative $i\tilde{\nabla}_\mu$ which includes spin connection 
with contorsions generated by gravitino\cite{GUSYNIN,OBUKHOV}. Then we obtain 
the supersymmetric version of Dirac operator, $\mathcal{D}_M^{SG}$ on the curved 
space-time. We investigate the square of the Dirac operator $P=(i\mathcal{D}_M^{SG})^2$ 
and the Seeley-Dewitt coefficients due to $P$ in order to obtain the action of 
supergravity. 
\section{Supersymmetrically extended triple on the flat space-time}
In our previous papers
, we introduced the triple for the supersymmetric theory which was extended from the spectral triple of the NCG on the 
flat Riemannian manifold. In this section, let us review it.
The functional space $\mathcal{H}_M$ on the Minkowskian space-time manifold 
is the direct sum of two subsets, $\mathcal{H}_+$ and 
$\mathcal{H}_-$:
\begin{equation}
\mathcal{H}_M = \mathcal{H}_+ \oplus \mathcal{H}_-. \label{HM}
\end{equation}
The element of $\mathcal{H}_M$ is given by
\begin{align}
\Psi  & = \begin{pmatrix}
\Psi_+ \\
\Psi_-
\end{pmatrix}
=\Phi_+ + \Phi_-, 
\label{Phi} \\
& \Phi_+=
\begin{pmatrix}
\Psi_+ \\
0^3 
\end{pmatrix} \in \mathcal{H}_+,\ 
\Phi_-= \begin{pmatrix}
0^3 \\
\Psi_-
\end{pmatrix}
\in \mathcal{H}_-.  \label{HpHm}
\end{align} 
Here, $\Psi_+$, $\Psi_-$ are denoted by 
\begin{equation}
(\Psi_+)_i = (\varphi_+(x),\psi_{+\alpha}(x),F_+(x))^T,\ i=1,2,3,  \label{Psi+}
\end{equation}
and 
\begin{equation}
(\Psi_-)_{\bar{i}} = (\varphi_-(x),\bar{\psi}_-^{\dot{\alpha}}(x),F_-(x))^T,\ \bar{i}=1,2,3, \label{Psi-}
\end{equation}
in the vector notation. Here, $\varphi_+$ and $F_+$ of $\Psi_+$ are complex scalar functions with mass dimension one and 
two, respectively, and $\psi_{+ \alpha}$,~$\alpha=1,2$ are 
the Weyl spinors on the space-time $M$ which 
have mass dimension $\frac{3}{2}$  and transform as the $(\frac{1}{2},0)$ representation of 
the Lorentz group, $SL(2,C)$. 
$\Psi_+(x)$ obey the following chiral supersymmetry transformation and 
form a chiral supermultiplet.
\begin{equation}
\begin{cases}
\delta_\xi\varphi_+ =\sqrt{2}\xi^\alpha \psi_{+\alpha}, \\
\delta_\xi\psi_{+\alpha} = {\rm i}\sqrt{2}\sigma^\mu_{\alpha\dot{\alpha}}\bar{\xi}^{\dot{\alpha}}\partial_\mu\varphi_+
+\sqrt{2}\xi_\alpha F_+,\\
\delta_\xi F_+ = {\rm i}\sqrt{2}\bar{\xi}_{\dot{\alpha}}\bar{\sigma}^{\mu\dot{\alpha}\alpha}\partial_\mu\psi_{+\alpha}.
\end{cases} \label{deltaxi}
\end{equation}

On the other hand, $\bar{\psi}^{\dot{\A}}$ transform as 
the $(0,\frac{1}{2})$ of $SL(2,C)$ and $\Psi_-(x)$ form an
 antichiral supermultiplet which obey the 
antichiral supersymmetry transformation as follows:
\begin{equation}
\begin{cases}
\delta_\xi\varphi_- = \sqrt{2}\bar{\xi}_{\dot{\alpha}}\bar{\psi}^{\dot{\alpha}}_-, \\
\delta_\xi\bar{\psi}_-^{\dot{\alpha}} = {\rm i}\sqrt{2}\bar{\sigma}^{\mu\dot{\alpha}\alpha}\xi_\alpha\partial_\mu\varphi_-
+\sqrt{2}\bar{\xi}^{\dot{\alpha}}F_-, \\
\delta_\xi F_- = {\rm i}\sqrt{2}\xi^\alpha\sigma^\mu_{\alpha\dot{\alpha}}\partial_\mu\bar{\psi}_-^{\dot{\alpha}}. 
\end{cases} \label{deltabxi}
\end{equation}

The Z/2 grading of the functional space $\mathcal{H}_M$ is given by an operator which is 
defined by
\begin{equation}
\gamma_M =\begin{pmatrix}
-{\rm i} & 0\\
0 & {\rm i}
\end{pmatrix}.
\end{equation}
In this basis, we have $\gamma_M(\Psi_+)=-{\rm i}$ and $\gamma_M(\Psi_-)={\rm i}$. 
Hereafter, we suitably abbreviate unit matrices or subscripts which denote sizes of unit and zero matrices. 

For the state $\Psi \in \mathcal{H}_M$ , the charge conjugate state $\Psi^c$ is given by
\begin{equation}
\Psi^c = \begin{pmatrix}
\Psi_+^c\\
\Psi_-^c
\end{pmatrix}.
\end{equation}
The antilinear operator $\mathcal{J}_M$ is defined by
\begin{equation}
\Psi^c = \mathcal{J}_M\Psi =C\Psi^*,
\end{equation}
so that it is given by
\begin{equation}
\mathcal{J}_M = C\otimes *,
\end{equation}
where C is the following charge conjugation matrix:
\begin{equation}
C =\left(\begin{array}{ccc|ccc}
 & & & 1 & 0 & 0  \\
 & \mathbf{0} & & 0 & \epsilon_{\alpha\beta} & 0 \\
   & & & 0 & 0 & 1  \\
  \hline 
  1 & 0 & 0 & & &  \\
  0 & \epsilon^{\dot{\alpha}\dot{\beta}} & 0 & & \mathbf{0} & \\
  0 & 0 & 1 & & &
\end{array}\right),
\end{equation}
and $*$ is the complex conjugation. The operator $\mathcal{J}_M $ obeys the 
following relation:
\begin{equation} 
\mathcal{J}_M\gamma_M =\gamma_M\mathcal{J}_M.
\end{equation}
The real structure $J_M$ is now expressed for the basis 
of the Hilbert space $(\Phi,\Phi^c)^T$ in the following form:
\begin{equation}
J_M = \begin{pmatrix}
0 & \mathcal{J}_M^{-1}\\
\mathcal{J}_M & 0
\end{pmatrix}. 
\end{equation}
The Z/2 grading $\Gamma_M$ on the basis is expressed by
\begin{equation}
\Gamma_M = \begin{pmatrix}
\gamma_M & 0\\
0 & \gamma_M
\end{pmatrix}.
\end{equation}

Corresponding to the construction of the functional space (\ref{HM}), 
the algebra $\mathcal{A}$ represented by them are expressed as
\begin{align}
\mathcal{A}_M & = \mathcal{A}_+ \oplus \mathcal{A}_-.
\end{align}
Here an element $u_a$ of $\mathcal{A_+}$, which acts on
 $\mathcal{H}_+$,
 and an element $\bar{u}_a$ of 
$\mathcal{A}_-$, which acts on $\mathcal{H}_-$ are given by 
\begin{align}
(u_a)_{ij} & = \frac{1}{m_0}
\begin{pmatrix}
\varphi_a & 0 & 0\\
\psi_{a\alpha} & \varphi_a & 0 \\
F_a & -\psi_a^\alpha & \varphi_a
\end{pmatrix} \in \mathcal{A}_+, 
\label{ua}
\\
(\bar{u}_a)_{\bar{i}\bar{j}} & = \frac{1}{m_0}
\begin{pmatrix}
\varphi_a^\ast & 0 & 0\\
\bar{\psi}_a^{\dot{\alpha}} & \varphi_a^\ast & 0 \\
 F_a^\ast & -\bar{\psi}_{a\dot{\alpha}} & \varphi_a^\ast
\end{pmatrix} \in \mathcal{A}_-, \label{barua},
\end{align}
where $\{\varphi_a(\varphi_a^\ast), \psi_{a\alpha}(\bar{\psi}_a^{\dot{\alpha}}),F_a(F_a^\ast)\}$
are chiral(antichiral) multiplets.

Note that these multiplets are not related to the multiplets
in the functional space in Eqs.~(\ref{Psi+}) and (\ref{Psi-}).
The elements of $\mathcal{A}$, 
$u_a$ and $\bar{u}_a$ together with the Dirac operator are 
the origin of the gauge and Higgs supermultiplets, while the 
elements (\ref{Psi+}) and (\ref{Psi-})
of the functional space are the origin of matter fields.

On the basis $(\Phi,\Phi^c)^T$, the Dirac operator $D_M$ on the manifold 
is given by 
\begin{equation}
D_M = \begin{pmatrix}
\mathcal{D}_M & 0\\
0 & \mathcal{J}_M \mathcal{D}_M \mathcal{J}_M^{-1}
\end{pmatrix},
\end{equation}
and
\begin{equation}
\mathcal{D}_M = -{\rm i}\begin{pmatrix}
0 & \bar{\mathcal{D}}_{i\bar{j}} \\
\mathcal{D}_{\bar{i}j} & 0
\end{pmatrix}, 
\label{DM0}
\end{equation}
where
\begin{equation}
\mathcal{D}_{\bar{i}j}=\begin{pmatrix}
0 & 0 & 1\\
0 & {\rm i}\bar{\sigma}^\mu\partial_\mu & 0\\
\Box & 0 & 0
\end{pmatrix},\ 
\bar{\mathcal{D}}_{i\bar{j}}= \begin{pmatrix}
0 & 0 & 1\\
0 & {\rm i}\sigma^\mu\partial_\mu & 0\\
\Box & 0 & 0
\end{pmatrix}.
\label{DM}
\end{equation}
We verified in Ref.\cite{paper0} that the Dirac operator and the supersymmetric transformation expressed by 
Eq.(\ref{deltaxi}) and (\ref{deltabxi}) were commutative. 

When we change the order of elements in the basis (\ref{Phi}),(\ref{HpHm}),(\ref{Psi+}),(\ref{Psi-}) to 
\begin{equation}
(\varphi_+,\varphi_-, \psi_{+\alpha},\psi_-^{\dot{\alpha}},F_+,F_-)\in \mathcal{H}_M \label{base1},
\end{equation}
the Dirac operator (\ref{DM0}) is replaced with 
\begin{equation}
\mathcal{D}_M = -i
\begin{pmatrix}
0 & 0 & 1_2 \\
0 & i\gamma^\mu\partial_\mu & 0\\
\Box\times 1_2 & 0 & 0
\end{pmatrix}. \label{DMflat}
\end{equation} 
We note again that the above formalism was given in the framework of Minkowskian signature 
in order to incorporate supersymmetry. 
When we restrict the functional space $\mathcal{H}_M$ to its fermionic part $\mathcal{H}_0$ and 
transfer to the Euclidean signature, 
we recover the original spectral triple which gives the framework of NCG.  
\section{Dirac operator on the curved space-time}
In order to obtain the supersymmetric Dirac operator on a curved space-time, 
we must consider torsion tensor.  
The torsion consists of gravitino $\psi_\mu$ which is a majorana spinor vector.   
\begin{equation}
T_{\lambda\mu\nu}=-\frac{1}{2}\ovl{\psi}_\mu\gamma_\lambda\psi_\nu=\frac{1}{2}\ovl{\psi}_\nu\gamma_\lambda\psi_\mu
\end{equation}
It is the antisymmetric part of affine connection expressed by 
\begin{equation}
T^\lambda_{\ \mu\nu}=\tilde{\Gamma}^\lambda_{\mu\nu}-\tilde{\Gamma}^\lambda_{\nu\mu}.
\end{equation}
The affine connection with the torsion is a sum of Christoffel symbol $\Gamma^\lambda_{\mu\nu}$ 
and contorsion $Y^\lambda_{\ \mu\nu}$:
\begin{equation}
\tilde{\Gamma}^\lambda_{\mu\nu}=\Gamma^\lambda_{\mu\nu}+Y^\lambda_{\ \mu\nu}.
\end{equation} 
The relation between the contorsion and the torsion is given by
\begin{equation}
Y_{\lambda\mu\nu} =\frac{1}{2}(T_{\lambda\mu\nu}+T_{\mu\nu\lambda}+T_{\nu\mu\lambda}).
\end{equation} 
The spin connection is separated to the contorsion and the part without contorsion\cite{Lopez}:
\begin{align}
\tilde{\omega}^{ab}_{\ \ \mu} & =\omega^{ab}_{\ \ \mu}+Y^{ab}_{\ \ \mu}, \\
Y^{ab}_{\ \ \mu} & = e^a_\rho e^b_\sigma Y^{\rho\sigma}_{\ \ \ \mu},
\end{align} 
where $e^a_\mu$ is vielbein which connects general coordinates denoted by subscript of the Greek letter 
to local inertial coordinates denoted by that of Roman letter. 
The covariant derivative for a spinor in the curved space is described by
\begin{equation}
\tilde{\nabla}_\mu =\partial_\mu+\tilde{\omega}_\mu, \label{coderivative1}
\end{equation}  
where  $\tilde{\omega}_\mu$ is a sum of products of the spin connection and commutator of $\gamma$ matrices 
$\gamma_{ab}=\frac{1}{2}[\gamma_a,\gamma_b] $ which is expressed by
\begin{align}
\tilde{\omega}_\mu & = \frac{1}{4}\tilde{\omega}^{ab}_{\ \ \mu}\gamma_{ab}
=\frac{1}{4}(\omega^{ab}_{\ \ \mu}+Y^{ab}_{\ \ \mu})\gamma_{ab}.
\end{align}

On the curved space, for the (2,2)-th entry of the matrix (\ref{DMflat}), we 
replace the partial derivative $\partial_\mu$ with the covariant derivative and for the (1,3)-th and (3,1)-th 
entries which act on bosonic wave functions, we adopt the operator which appears in 
the equation given by the action of the Klein-Gordon field in the curved space\cite{Fiorenzo}. 
So, the Dirac operator on the curved space 
is expressed by
\begin{align}
\mathcal{D}^{(SG)}_M = & -i
\begin{pmatrix}
0 & 0 & 1_2\\
0 & i\gamma^\mu\tilde{\nabla}_\mu & 0\\
(g^{\mu\nu}\tilde{\nabla}_\mu\tilde{\nabla}_\nu+\upsilon \tilde{R})\times 1_2 & 0 & 0
\end{pmatrix} \nonumber \\
& =
-i
\begin{pmatrix}
0 & 0 & 1_2\\
0 & i\gamma^a e_a^\mu(\partial_\mu+\tilde{\omega}_\mu) & 0\\
\left(g^{\mu\nu}(\partial_\mu\partial_\nu- \tilde{\Gamma}^\rho_{\mu\nu}\partial_\rho)
+\upsilon\tilde{R}\right)\times 1_2 & 0 & 0
\end{pmatrix}, \label{DMSG}
\end{align} 
where $\tilde{R}$ is curvature with the torsion and $\upsilon$ is an unknown constant. 
\section{Supergravity action}
In our noncommutative geometric approach to supersymmetry, the action for supergravity will be obtained 
by the coefficients of heat kernel expansion of the 
operator $P=(\mathcal{D}^{(SG)}_M)^2$\cite{gilkey}. The prescription to obtain these coefficients 
on the curved space with the torsion 
for the spinorial part of $\mathcal{D}^{(SG)}_M$, i.e. (2,2)-th entry of the matrix (\ref{DMSG})  
and its result are given by \cite{OBUKHOV}. 

We want to obtain the coefficients for $\mathcal{D}^{(SG)}_M$ including (1,3)-th and (3,1)-th entries. 
At first, we expand the operator $P$ into the following form:
\begin{equation}
P = -(g^{\mu\nu}\tilde{\nabla}_\mu\tilde{\nabla}_\nu +\tilde{\mathbb{A}}^\mu\tilde{\nabla}_\mu +\tilde{\mathbb{B}}). \label{P} 
\end{equation}  
We define a vector $S_\mu$ as follows:
\begin{equation}
S_\mu = Q_\mu+\tilde{\mathbb{A}}_\mu, \label{Smu}
\end{equation}
where $Q_\mu$ is torsion trace $T^\alpha_{\ \mu\alpha}$. 
In our theory, trace over the vector bundle 
are replaced with supertrace. 
When a matrix $M$ in the basis (\ref{base1}) is given by
\begin{equation}
M=\begin{pmatrix}
M_{11} &  M_{12} & M_{13}\\
M_{21} & M_{22} & M_{23}\\
M_{31} & M_{32} & M_{33}
\end{pmatrix},
\end{equation}
the supertrace is expressed by
\begin{equation}
{\rm Str} M = {\rm tr_V} M_{11} +{\rm tr_V}M_{33}-{\rm tr_V} M_{22}. 
\end{equation} 

Since the bosonic degrees of freedom equals that of fermionic states, the supertrace of $\mathbb{I}$ vanishes. 
Then the coefficients $a_n(P)$ are given by
\begin{align} 
a_0(P) & =\frac{1}{16\pi^2}\int_M
{\rm d}^4x\sqrt{-g}
 {\rm Str}(\mathbb{I})=0, \label{a0} \\
a_2(P) & =\frac{1}{16\pi^2}\int_M {\rm d}^4x \sqrt{-g}
{\rm Str}(\frac{1}{6}R\mathbb{I}+\mathbb{Z})
=\frac{1}{16\pi^2}\int_M {\rm d}^4x \sqrt{-g}{\rm Str} \mathbb{Z}, \label{a2} \\
a_4(P) & =\frac{1}{16\pi^2}\int_M {\rm d}^4x \sqrt{-g}
\frac{1}{360}{\rm Str}\left((
12\Box R+5R^2-2R_{\mu\nu}R^{\mu\nu}+2R_{\mu\nu\rho\sigma}R^{\mu\nu\rho\sigma})\mathbb{I}
\right. \nonumber \\
& \left. +60R\mathbb{Z}+180\mathbb{Z}^2+60\Box\mathbb{Z}+30\Omega_{\mu\nu}\Omega^{\mu\nu}
\right) \nonumber \\
= & \frac{1}{16\pi^2}
\int_M {\rm d}^4x\sqrt{-g}
\frac{1}{360}{\rm Str}(60R\mathbb{Z}+180\mathbb{Z}^2+60\Box\mathbb{Z}+30\Omega_{\mu\nu}\Omega^{\mu\nu}),
\label{a4}
\end{align}
where $\Omega^{\mu\nu}$ is the bundle curvature that we will describe later 
and $\mathbb{Z}$ is a function defined as follows:
\begin{align}
\mathbb{Z} & = \tilde{\mathbb{B}} -\frac{1}{2}\nabla_\mu S^\mu +\frac{1}{4}S^\mu S_\mu. 
\end{align}

After some algebra in terms of the Riemann curvature tensor $\tilde{R}^\lambda_{\ \rho\mu\nu}$ with torsion in appendix A, 
the square of (\ref{DMSG}) is given by
\begin{equation}
P=\mathcal{D}_M^{(SG)2} =  -\begin{pmatrix}
D_2^{(\varphi)} & 0 & 0\\
0 & D_2^{(\psi)} & 0\\
0 & 0 & D_2^{(\varphi)}
\end{pmatrix} , \label{D_2}
\end{equation}
where 
\begin{align}
D_2^{(\varphi)} = & g^{\mu\nu}\tilde{\nabla}_\mu\tilde{\nabla}_\nu + \upsilon\tilde{R}, \\
D_2^{(\psi)} = & g^{\mu\nu}\tilde{\nabla}_\mu\tilde{\nabla}_\nu
 -\frac{1}{2}\gamma^{\mu\nu}T^\lambda_{\ \mu\nu}\tilde{\nabla}_\lambda
 +\frac{1}{8}\gamma^{\mu\nu}\gamma_{ab}e_\lambda^a e^{\rho b}\tilde{R}^\lambda_{\ \rho\mu\nu}. \label{D2PSI}
\end{align}
On the basis (\ref{base1}), the functions $\tilde{\mathbb{A}},\tilde{\mathbb{B}}$ in (\ref{P}) and $\mathbb{S}_\mu$ in (\ref{Smu}) 
are expressed by the matrix form as follows:
\begin{equation}
\tilde{\mathbb{A}}^\mu = \begin{pmatrix}
0 & 0 & 0\\
0 & \tilde{\mathbb{A}}^\mu_{(\psi)} & 0\\
0 & 0 & 0
\end{pmatrix} , 
\tilde{\mathbb{B}} = \begin{pmatrix}
\upsilon \tilde{R} & 0 & 0\\
0 & \tilde{\mathbb{B}}_{(\psi)} & 0\\
0 & 0 & \upsilon\tilde{R}
\end{pmatrix},
S_\mu = \begin{pmatrix}
Q_\mu & 0 & 0\\
0 & S_\mu^{(\psi)} & 0\\
0 & 0 & Q_\mu
\end{pmatrix}, \label{ABSmu}
\end{equation}
where 
\begin{equation}
\tilde{\mathbb{A}}^\mu_{(\psi)} =  -\frac{1}{2}\gamma^{\alpha\beta}T^\mu_{\ \alpha\beta},\ \ 
\tilde{\mathbb{B}}_{(\psi)} =  \frac{1}{8}\gamma^{\mu\nu}\gamma_{ab}e_\lambda^a e^{\rho b}\tilde{R}^\lambda_{\ \rho\mu\nu},\ \ 
S_\mu^{(\psi)} = Q_\mu -\frac{1}{2}\gamma^{\alpha\beta}T^\mu_{\ \alpha\beta}.
\end{equation}
Then the matrix form of the function $\mathbb{Z}$ is given by
\begin{equation}
\mathbb{Z} =
\begin{pmatrix}
Z^{(\varphi)}\times 1_2 & 0 & 0\\
0 & Z^{(\psi)} & 0 \\
0 & 0 & Z^{(\varphi)}\times 1_2
\end{pmatrix}, \label{matrixZ}
\end{equation}
where 
\begin{align}
Z^{(\varphi)}  = & g^{\mu\nu}(-\frac{1}{2}\nabla_\mu T^\alpha_{\ \nu\alpha} -\frac{1}{4}T^\alpha_{\ \mu\alpha}T^\beta_{\ \nu\beta})+\upsilon \tilde{R}
= g^{\mu\nu}(-\frac{1}{2} \tilde{\nabla}_\mu Q_\nu+\frac{1}{4}Q_\mu Q_\nu)+\upsilon\tilde{R},
\label{Zphi} \\
Z^{(\psi)} = & \frac{1}{8}\gamma^{\mu\nu}\gamma^{\lambda\rho}\tilde{R}_{\lambda\rho\mu\nu}
-\frac{1}{2}\nabla_\mu S^{(\psi)\mu}-\frac{1}{4}S^{(\psi)\mu} S^{(\psi)}_\mu \nonumber \\
= & \frac{1}{8}\gamma^{\mu\nu}\gamma^{\lambda\rho} \tilde{R}_{\lambda\rho\mu\nu} 
-\frac{1}{2}\tilde{\nabla}_\mu(Q^\mu -\frac{1}{2}\gamma^{\alpha\beta}T^\mu_{\ \ \alpha\beta}) 
+\frac{1}{4}Q_\mu Q^\mu 
-\frac{1}{16}\gamma^{\alpha\beta}\gamma^{\rho\sigma}T_{\mu\alpha\beta}T^\mu_{\ \rho\sigma}. \label{Zpsi}
\end{align} 

Using Eq.(\ref{Zphi}) and Eq.(\ref{Zpsi}), $a_2(P)$ of (\ref{a2}) can be converted into  
\begin{align}
a_2(P) = & \frac{1}{16\pi^2}\int_M {\rm d}^4x\sqrt{-g}(4 Z^{(\varphi)}-Tr Z^{(\psi)}) 
= \frac{1}{16\pi^2}\int_M {\rm d}^4x \sqrt{-g}\left((4\upsilon+1)\tilde{R} 
- \frac{1}{2}T^{\mu\rho\sigma}T_{\mu\rho\sigma} \right) \nonumber \\
 = & \frac{1}{16\pi^2}\int_M {\rm d}^4x \sqrt{-g}
\left(
(4\upsilon+1)(R+\frac{1}{2}T^{\mu\nu\lambda}T_{\lambda\nu\mu}-Q_\mu Q^\mu -2\nabla_\mu Q^\mu) +2\upsilon T^{\mu\nu\lambda} T_{\mu\nu\lambda}\right).
\label{a2lambda}
\end{align}
 
In the basis (\ref{base1}), the bundle curvature $\Omega^{\mu\nu}$ in Eq.(\ref{a4}) is also given by 
\begin{equation}
\Omega_{\mu\nu} =
\begin{pmatrix}
\Omega_{\mu\nu}^{(\varphi)}\times 1_2 & 0 & 0\\
0 & \Omega_{\mu\nu}^{(\psi)} & 0\\
0 & 0 & \Omega_{\mu\nu}^{(\varphi)}\times 1_2
\end{pmatrix}, \label{Omega}
\end{equation}
where $\Omega_{\mu\nu}^{(\varphi)}$ is given by 
\begin{equation}
\Omega_{\mu\nu}^{(\varphi)} =  
\frac{1}{2}((\nabla_\mu Q_\nu)-(\nabla_\nu Q_\mu))
= 
\frac{1}{2}((\tilde{\nabla}_\mu Q_\nu) -(\tilde{\nabla}_\nu Q_\mu)) +\frac{1}{2}T^\lambda_{\ \nu\mu}Q_\lambda, 
\label{Omegaphi}
\end{equation}
and using (\ref{tildeR}), $\Omega_{\mu\nu}^{(\psi)}$ is also obtained by
\begin{align} 
\Omega_{\mu\nu}^{(\psi)} = &
\partial_\mu\tilde{\omega}_\nu-\partial_\nu\tilde{\omega}_\mu+
[\tilde{\omega}_\mu,\tilde{\omega}_\nu]+\frac{1}{2}(\nabla_\mu S^{(\psi)}_\nu-\nabla_\nu S^{(\psi)}_\mu)
+\frac{1}{4}[S_\mu^{(\psi)},S^{(\psi)}_\nu] \nonumber \\
= & -\frac{1}{4}\gamma^{\lambda\rho}\tilde{R}_{\lambda\rho\mu\nu}
+\frac{1}{2}(\tilde{\nabla}_\mu S^{(\psi)}_\nu-\tilde{\nabla}_\nu S^{(\psi)}_\mu)
+\frac{1}{4}[S_\mu^{(\psi)},S^{(\psi)}_\nu]-\frac{1}{2}T^\lambda_{\ \mu\nu}S_\lambda \label{Omegapsi}.
\end{align}

After long and tedious algebra using the supertraces in appendix B, the coefficient $a_4(P)$ of (\ref{a4}) is 
converted into
\begin{equation}
 a_4(P) = \frac{1}{16\pi^2}\int_M {\rm d}^4x\sqrt{-g}((\partial)^{(2)} + R^{(2)}+R\partial T+RT^{(2)}
+(\partial T)^{(2)}+T^{(2)}\partial T +T^{(4)}),
\label{a4lambda}
\end{equation}
where
\begin{align}
(\partial)^{(2)} = & \frac{1}{6}\Box \left((1+4\upsilon)\tilde{R}
-\frac{1}{2}T^{\mu\rho\sigma}T_{\mu\rho\sigma}\right), \\
R^{(2)} =& (2\upsilon^2+\frac{2}{3}\upsilon+\frac{1}{24})\tilde{R}^2+\frac{1}{24}\tilde{R}_{\mu\nu\lambda\rho}\tilde{R}^{\mu\nu\lambda\rho}, \\
R\partial T = & -\frac{1}{6}(1+4\upsilon)\tilde{R}\tilde{\nabla}_\alpha Q^\alpha
+\frac{1}{3}\tilde{R}_{\mu\nu\lambda\rho}\tilde{\nabla}^\lambda T^{\rho\mu\nu}, \\
RT^{(2)} = & \frac{1}{12}(1+4\upsilon)\tilde{R}Q_\alpha Q^\alpha -\frac{1}{6}\upsilon\tilde{R}T_{\mu\nu\lambda}T^{\mu\nu\lambda}
-\frac{1}{12}(1+4\upsilon )\tilde{R}T^{\mu\nu\lambda}T_{\lambda\nu\mu} \nonumber \\
& - \frac{2}{3}\tilde{R}_{\mu\nu\lambda\rho}T^{\lambda\sigma\mu}T^{\rho\ \nu}_{\ \sigma} 
-\frac{1}{2}\tilde{R}_{\mu\nu\lambda\rho}T_\sigma^{\ \mu\lambda}T^{\sigma\nu\rho} 
+\frac{1}{6}\tilde{R}_{\mu\nu\lambda\rho}T_\sigma^{\ \lambda\rho}T^{\sigma\mu\nu}, \\
(\partial T)^{(2)} = & -\frac{1}{48}(\tilde{\nabla}_\mu T_{\rho\alpha\beta})(\tilde{\nabla}^\rho T^{\mu\alpha\beta})
+ \frac{1}{2}((\tilde{\nabla}_\mu Q_\nu)(\tilde{\nabla}^\mu Q^\nu) 
-(\tilde{\nabla}_\mu Q_\nu)(\tilde{\nabla}^\nu Q^\mu)) \nonumber \\
& +\frac{1}{48}(\tilde{\nabla}_\mu T_{\rho\alpha\beta})(\tilde{\nabla}^\mu T^{\rho\alpha\beta}) 
+ \frac{1}{8}(-(\tilde{\nabla}_\mu T_{\rho\alpha\beta})(\tilde{\nabla}^\alpha T^{\rho\beta\mu})
- (\tilde{\nabla}_\mu T_{\rho\alpha\beta})(\tilde{\nabla}^\alpha T^{\beta\mu\rho}) \nonumber \\
& -(\tilde{\nabla}_\mu T_{\rho\alpha\beta})(\tilde{\nabla}^\mu T^{\beta\rho\alpha})  
- (\tilde{\nabla}_\mu T_{\rho\alpha\beta})(\tilde{\nabla}^\rho T^{\beta\alpha\mu})  
+(\tilde{\nabla}_\mu T_{\rho\alpha\beta})(\tilde{\nabla}^\alpha T^{\mu\beta\rho}) 
) \\
T^{(2)}\partial T = & \frac{1}{2}(\tilde{\nabla}_\mu Q^\nu - \tilde{\nabla}_\nu Q_\mu)Q_\alpha T^{\alpha\mu\nu} 
-\frac{5}{12}(\tilde{\nabla}_\alpha Q^\alpha)T_{\lambda\mu\nu}T^{\lambda\mu\nu} 
- (\tilde{\nabla}_\mu T_\nu^{\ \alpha\beta})(
\frac{1}{4}T^{\sigma\mu}_{\ \ \ \alpha}T_{\beta\sigma}^{\ \ \ \nu} \nonumber \\
& + \frac{1}{4}T^{\sigma\nu}_{\ \ \ \alpha}T_{\beta\ \sigma}^{\ \mu}
+ \frac{1}{4}T^{\sigma\nu\mu}T_{\alpha\beta\sigma}-\frac{1}{8}T^{\mu\nu\sigma}T_{\sigma\alpha\beta}
-\frac{1}{4}T^{\sigma\nu}_{\ \ \ \alpha}T^\mu_{\ \beta\sigma} 
+ \frac{1}{4}T^{\sigma\mu}_{\ \ \ \alpha}T^\nu_{\ \beta\sigma} \nonumber \\
& -\frac{3}{8}T_{\sigma\alpha\beta}T^{\nu\mu\sigma} 
-\frac{4}{3}T^{\mu\sigma}_{\ \ \ \alpha}T^\nu_{\ \sigma\beta}
- \frac{1}{6}T^{\sigma\mu\nu}T_{\sigma\alpha\beta}),
\end{align}
\begin{align}
T^{(4)}
= & -\frac{1}{24}Q_\alpha Q^\alpha T_{\mu\nu\lambda}T^{\mu\nu\lambda}
+\frac{1}{4}Q_\alpha T^\alpha_{\ \mu\nu}Q_\beta T^{\beta\mu\nu} \nonumber \\
+ & \frac{1}{8}T^{\xi\mu\nu}T_{\alpha\xi\beta}(T^{\zeta\ \beta}_{\ \nu}T_{\mu\zeta}^{\ \ \ \alpha}
+T^{\zeta\alpha\beta}T_{\mu\nu\zeta}
+T^{\zeta\alpha}_{\ \ \ \nu}T_{\mu\zeta}^{\ \ \ \beta} 
-T^{\zeta\ \alpha}_{\ \mu}T_{\ \zeta\nu}^{\beta}
-T^{\zeta\ \beta}_{\ \mu}T_{\ \zeta\nu}^{\alpha}
-T_{\zeta\mu\nu}T^{\alpha\zeta\beta})  \nonumber \\
- & \frac{1}{16}T^{\xi\mu\nu}T_{\alpha\xi\beta}(T^\zeta_{\ \nu\mu}T_{\beta\zeta\alpha}
+T^\zeta_{\ \nu\mu}T^{\alpha\ \beta}_{\ \zeta})
-\frac{1}{4}T^{\xi\mu\nu}T_{\alpha\xi\beta}T^{\zeta\beta\nu}T^\alpha_{\ \mu\zeta}
- \frac{1}{24}T^{\xi\mu\nu}T_{\xi\mu\nu}T_{\zeta\alpha\beta}T^{\zeta\alpha\beta} \nonumber \\
- & \frac{1}{6}T^{\xi\mu\nu}T_{\xi\alpha\beta}T_{\zeta\mu\nu}T^{\zeta\alpha\beta} 
-\frac{1}{4}T^{\xi\mu\nu}T_{\xi\alpha\beta}T_{\zeta\mu}^{\ \ \alpha}T^{\zeta\beta}_{\ \ \nu}
+\frac{1}{48}T^{\xi\mu\nu}T_{\xi\alpha\nu}T_{\zeta\mu\sigma}T^{\zeta\alpha\sigma} \nonumber\\
- & \frac{1}{48}T^{\xi\mu\nu}T_{\xi\alpha\beta}T_{\zeta\mu}^{\ \ \beta}T^{\zeta\alpha}_{\ \ \ \nu} 
-\frac{2}{3}T^{\xi\mu\nu}T_{\xi\alpha\beta}T_{\mu\zeta}^{\ \ \ \alpha}T_\nu^{\ \zeta\beta} 
+\frac{1}{24}T^{\xi\mu\nu}T_{\xi\mu\nu}T^{\zeta\alpha\beta}T_{\beta\alpha\zeta}.
\end{align}

In general, the coefficients $a_n(P)$ vanish for odd values of $n$\cite{gilkey}. In addition, the mass 
dimension of each term of the integrand in $a_n(P)$ is $n$ so that in order to conserve 
the renormalizabilty, $a_n(n>4)$  should not appear in the action. 
Then, we have obtained the all renormalizable terms of our supergravity action.   
\section{Conclusions and discussions}
In this paper, we have derived in Eq.(\ref{DMSG}) the supersymmtric Dirac operator $D_M^{(SG)}$  
on the Riemann-Cartan curved space without gauge interaction by replacing the derivative with respect to 
the space-time coordinates in Eq.(\ref{DMflat}) with the covariant derivative of the general coordinate transformation. 
This operator includes the spin connection, 
the affine connection and the curvature with torsion tensors which consist of gravitinos. According to the 
prescription of the spectral action principle, we have obtained the square of $D_M^{(SG)}$ and 
have taken it to pieces as Eq.(\ref{P}). 
We have replaced the trace in the ordinary spectral action with the supertrace and calculated the Seeley-Dewitt coefficients  
of the heat kernel expansion. The coefficient $a_0(P)$ in Eq.(\ref{a0}) cancels out. 
It means that the cosmological constant vanishes in the supersymmetric theory. 
So the supergravity action of our theory is given by
\begin{equation}
S = \alpha a_2(P) + \beta a_4(P), \label{sugraaction}
\end{equation}
where $\alpha$ and $\beta$ are some constants, the coefficients $a_2(P)$ and $a_4(P)$ are given 
in Eq.(\ref{a2lambda}) and (\ref{a4lambda}). It is a modified Einstein-Hilbert action. 

In the action (\ref{sugraaction}), there is no term with the Ricci curvature tensor. Therefore, 
when we construct based on NCG a gravity theory which possesses physically important property such as 
conformal invariance, renormalizability, if we want the theory able to be extended supersymmetrically, 
we should build its action not to include Ricci curvature tensor. 
For example, 
one of the simplest theory which possesses local conformal symmetry and renormalizabilty 
consists of the Weyl action term $\int {\rm d}^4 x \sqrt{-g} C^2 $ and the Gauss-Bonnet topological 
term and surface terms 
$\int {\rm d}^4 x \sqrt{-g}(\eta G_b + \tau \Box R) $ \cite{Berredo}
, where $C^2$ and $G_b$ terms are given by 
\begin{align}
C^2 = & R_{\mu\nu\lambda\rho}R^{\mu\nu\lambda\rho}-2R_{\mu\nu}R^{\mu\nu}+\frac{1}{3}R^2, \\
G_b = & R_{\mu\nu\lambda\rho}R^{\mu\nu\lambda\rho}-4R_{\mu\nu}R^{\mu\nu}+R^2.
\end{align}
The linear combination whose 
terms with Ricci tensor cancel out is given by
\begin{equation}
2C^2-G_b = R_{\mu\nu\lambda\rho}R^{\mu\nu\lambda\rho} -\frac{1}{3}R^2.
\end{equation} 
Seeing Eq.(\ref{tcurvet}) and Eq.(\ref{tcurve}), we know that in Eq.(\ref{a4lambda}), the coefficients of 
$\tilde{R}_{\mu\nu\lambda\rho}\tilde{R}^{\mu\nu\lambda\rho}$ and $\tilde{R}^2$ are same as those of 
$R_{\mu\nu\lambda\rho}R^{\mu\nu\lambda\rho}$ and $R^2$. 
So, let us take the ratio of them at $1:-\frac{1}{3}$ as follows:
\begin{equation}
\frac{1}{24} : 2\upsilon^2+\frac{2}{3}\upsilon+\frac{1}{24} = 1:-\frac{1}{3}.
\end{equation}
Then we obtain $\upsilon = -\frac{1}{6}$. The coefficient $a_2(P)$ in the Eq.(\ref{a2lambda}) is replaced with
\begin{align}
a_2(P) = & \frac{1}{16\pi^2}\int_M {\rm d}^4x\sqrt{-g}
(\frac{1}{3}\tilde{R}-\frac{1}{2}T^{\mu\rho\sigma}T_{\mu\rho\sigma}) \nonumber \\
= & \frac{1}{48\pi^2}\int_M {\rm d}^4x\sqrt{-g}
(R-2\nabla_\mu T^{\alpha\mu}_{\ \ \alpha}-T^{\alpha\mu}_{\ \ \ \alpha}T^\beta_{\ \mu\beta}
-\frac{5}{4}T^{\mu\nu\lambda}T_{\mu\nu\lambda}+\frac{1}{2}T^{\mu\nu\lambda}T_{\lambda\nu\mu}
). \label{a2R}
\end{align} 
The coefficient $a_4(P)$ is also replaced with  
\begin{align}
\lefteqn{a_4(P)}\nonumber \\
 = &  \frac{1}{16\pi^2}\int_M {\rm d}^4x\sqrt{-g}\left(
\frac{1}{6}\Box (\frac{1}{3}\tilde{R}-\frac{1}{2}T^{\mu\rho\sigma}T_{\mu\rho\sigma}) 
 +\frac{1}{24}(\tilde{R}_{\mu\nu\lambda\rho}\tilde{R}^{\mu\nu\lambda\rho}-\frac{1}{3}\tilde{R}^2)
-\frac{1}{18}\tilde{R}\tilde{\nabla}_\alpha Q^\alpha \right. \nonumber \\
& +\frac{1}{36}\tilde{R}Q_\alpha Q^\alpha +\frac{1}{36}\tilde{R}T_{\mu\nu\lambda}T^{\mu\nu\lambda}
-\frac{1}{36}\tilde{R}T^{\mu\nu\lambda}T_{\lambda\nu\mu} 
+\frac{1}{3}\tilde{R}_{\mu\nu\lambda\rho}\tilde{\nabla}^\lambda T^{\rho\mu\nu} \nonumber \\
& -\frac{2}{3}\tilde{R}_{\mu\nu\lambda\rho}T^{\lambda\sigma\mu}T^{\rho\ \nu}_{\ \sigma}
-\frac{1}{2}\tilde{R}_{\mu\nu\lambda\rho}T_\sigma^{\ \mu\lambda}T^{\sigma\nu\rho} 
+\frac{1}{6}\tilde{R}_{\mu\nu\lambda\rho}T_\sigma^{\ \lambda\rho}T^{\sigma\mu\nu} \nonumber \\
& \left. + (\partial T)^{(2)}+T^{(2)}\partial T + T^{(4)} \right). \label{a4R} 
\end{align}
When we reduce the coefficient $a_4(P)$ in Eq.(\ref{a4R}) to non-supersymmetric part,  
i.e. the part without terms including torsion tensor, 
it has captured the local conformal symmetry and renormalizability. Indeed, the coefficient $a_4(p)$  
includes a new type of non-supersymmetric gravity action 
$S_1$ which is given by 
\begin{equation}
S_1 = \frac{\beta}{364\pi^2}\int_M {\rm d}^4x\sqrt{-g}(R_{\mu\nu\lambda\rho}R^{\mu\nu\lambda\rho}
-\frac{1}{3}R^2). 
\end{equation} 
The variation of $S_1$ due to the conformal transformation $\delta g_{\mu\nu} = -\epsilon g_{\mu\nu}$ is given by 
\begin{equation}
\frac{1}{\sqrt{-g}}\delta S_1 = \frac{\beta\epsilon}{96\pi^2}\nabla_\mu\nabla_\nu G^{\mu\nu},
\end{equation}
where $G_{\mu\nu}$ is the Einstein tensor, $G^{\mu\nu}= R^{\mu\nu}-\frac{1}{2}g^{\mu\nu}R$.
Therefore, we can verify the conformal invariance of $S_1$ by the Bianchi's identity $\nabla_\mu G^{\mu\nu}=0$.

----------
\appendix{Appendix}
\renewcommand{\theequation}{A.\arabic{equation}}
\setcounter{equation}{0}
\section{Affine connection, spin connection, vielbein, curvature tensors with torsion}
In this appendix, we show some equations about affine connection, spin connection, vielbein, 
curvature tensors with torsion. 
The covariant derivatives of vielbein vanish. 
\begin{align}
\tilde{\nabla}_\mu e^a_\nu 
& =\partial_\mu e^a_\nu-\tilde{\omega}^a_{\ b\mu}e^b_\nu-\tilde{\Gamma}_{\ \nu\mu}^\lambda e^a_\lambda=0, \label{nablae1}\\
\tilde{\nabla}_\mu e_a^\nu & =\partial_\mu e_a^\nu-\tilde{\omega}_{a\ \mu}^{\ b}e_b^\nu+\tilde{\Gamma}^\nu_{\ \lambda\mu}e_a^\lambda
=0. \label{nablae2}
\end{align}
From (\ref{nablae1}), we obtain the relation between the affine connection and the spin connection expressed by
\begin{equation}
\tilde{\Gamma}_{\ \nu\mu}^\lambda = e^\lambda_a(\partial_\mu e^a_\nu-\tilde{\omega}^a_{\ b\mu}e^b_\nu). \label{Gamma}
\end{equation}
We also provide some else equations about  
the affine connection, the spin connection and the vielbein as follows:
\begin{align}
\partial_\mu\tilde{\Gamma}^\lambda_{\rho\nu} & =
(\partial_\mu e_a^\lambda)(\partial_\nu e_\rho^a-\tilde{\omega}^a_{\ b\nu}e_\rho^b)
+e_a^\lambda\left((\partial_\mu\partial_\nu e_\rho^a)-(\partial_\mu\tilde{\omega}^{ab}_{\ \ \nu})e_{b\rho}
-\tilde{\omega}^{ab}_{\ \ \nu}(\partial_\mu e_{b \rho})\right), \\
\partial_\nu\tilde{\Gamma}^\lambda_{\rho\mu} & =
(\partial_\nu e_a^\lambda)(\partial_\mu e_\rho^a-\tilde{\omega}^a_{\ b\mu}e_\rho^b)
+e_a^\lambda\left((\partial_\nu\partial_\mu e_\rho^a)-(\partial_\nu\tilde{\omega}^{ab}_{\ \ \mu})e_{b\rho}
-\tilde{\omega}^{ab}_{\ \ \mu}(\partial_\nu e_{b \rho})\right), \\
\tilde{\Gamma}^\lambda_{\sigma\mu}\tilde{\Gamma}^\sigma_{\rho\nu} & =
e_a^\lambda(\partial_\mu e_\sigma^a-\tilde{\omega}^{ab}_{\ \ \mu}e_{b\sigma})
e^\sigma_c(\partial_\nu e_\rho^c-\tilde{\omega}^{cd}_{\ \ \nu}e_{d\rho}), \nonumber \\
& = e_a^\lambda(\partial_\mu e_\sigma^a)e_c^\sigma\left((\partial_\nu e_\rho^c)-\tilde{\omega}^{cd}_{\ \ \nu}e_{d\rho}\right)
-e_a^\lambda \tilde{\omega}^{ab}_{\ \ \mu}\eta_{bc}\left((\partial_\nu e_\rho^c)-\tilde{\omega}^{cd}_{\ \ \nu}e_{d\rho}\right), \\
\tilde{\Gamma}^\lambda_{\sigma\nu}\tilde{\Gamma}^\sigma_{\rho\mu} 
& = e_a^\lambda(\partial_\nu e_\sigma^a)e_c^\sigma\left((\partial_\mu e_\rho^c)-\tilde{\omega}^{cd}_{\ \ \mu}e_{d\rho}\right)
-e_a^\lambda \tilde{\omega}^{ab}_{\ \ \nu}\eta_{bc}\left((\partial_\mu e_\rho^c)-\tilde{\omega}^{cd}_{\ \ \mu}e_{d\rho}\right). 
\label{GammaGamma}
\end{align}
Here, since 
\begin{equation}
e_c^\sigma(\partial_\nu e_\rho^c) = (\partial_\nu e_c^\sigma e_\rho^c)-(\partial_\nu e_c^\sigma)e_\rho^c =
(\partial_\nu \delta_\rho^\sigma)-(\partial_\nu e_c^\sigma)e_\rho^c=-(\partial_\nu e_c^\sigma)e_\rho^c, 
\end{equation}
we obtain one more equation as follows:
\begin{align}
e_a^\lambda(\partial_\mu e_\sigma^a)e_c^\sigma\left((\partial_\nu e_\rho^c)
-\tilde{\omega}^{cd}_{\ \ \nu}e_{d\rho}\right)
-e_a^\lambda(\partial_\nu e_\sigma^a)e_c^\sigma\left((\partial_\mu e_\rho^c)
-\tilde{\omega}^{cd}_{\ \ \mu}e_{d\rho}\right) & \nonumber \\
= -(\partial_\mu e_a^\lambda)
\left(
(\partial_\nu e_\rho^a)-\tilde{\omega}^{ad}_{\ \ \nu}e_{d\rho}
\right)
 +(\partial_\nu e_a^\lambda)\left((\partial_\mu e_\rho^a)
 -\tilde{\omega}^{ad}_{\ \ \mu}e_{d\rho}\right). & \label{epe}
\end{align}
Using Eq.(\ref{Gamma})-Eq.(\ref{epe}), we obtain the expression of the Riemann curvature tensor $\tilde{R}^\lambda_{\ \rho\mu\nu}$ 
with torsion as follows:
\begin{align}
\tilde{R}^\lambda_{\ \rho\mu\nu} & =
\partial_\mu\tilde{\Gamma}^\lambda_{\rho\nu}-\partial_\nu\tilde{\Gamma}^\lambda_{\rho\mu}
+\tilde{\Gamma}^\lambda_{\sigma\mu}\tilde{\Gamma}^\sigma_{\rho\nu}
-\tilde{\Gamma}^\lambda_{\sigma\nu}\tilde{\Gamma}^\sigma_{\rho\mu} \nonumber \\
= & (\partial_\mu e_a^\lambda)(\partial_\nu e_\rho^a)-(\partial_\nu e_a^\lambda)(\partial_\mu e_\rho^a)
-(\partial_\mu e_a^\lambda)\tilde{\omega}^{ab}_{\ \ \nu}e_{b\rho}
+(\partial_\nu e_a^\lambda)\tilde{\omega}^{ab}_{\ \ \mu}e_{b\rho} \nonumber \\
& - e_a^\lambda\left((\partial_\mu\tilde{\omega}^{ab}_{\ \ \nu})
-(\partial_\nu\tilde{\omega}^{ab}_{\ \ \mu})\right)e_{b\rho}
- e_a^\lambda\left(\tilde{\omega}^{ab}_{\ \ \nu}(\partial_\mu e_{b\rho})
-\tilde{\omega}^{ab}_{\ \ \mu}(\partial_\nu e_{b\rho})\right)\nonumber\\
& +e_a^\lambda(\partial_\mu e_\sigma^a)(e_c^\rho(\partial_\nu e_\rho^c)-\tilde{\omega}^{cd}_{\ \ \nu}e_{d\rho})
-e_a^\lambda \tilde{\omega}^{ab}_{\ \ \mu}\eta_{bc}\left((\partial_\nu e_\rho^c)-\tilde{\omega}^{cd}_{\ \ \nu}e_{d\rho}\right)
\nonumber \\
& -e_a^\lambda(\partial_\nu e_\sigma^a)e_c^\sigma\left((\partial_\mu e_\rho^c)-\tilde{\omega}^{cd}_{\ \ \mu}e_{d\rho}\right) 
+e_a^\lambda \tilde{\omega}^{ab}_{\ \ \nu}\eta_{bc}\left((\partial_\mu e_\rho^c)-\tilde{\omega}^{cd}_{\ \ \mu}e_{d \rho}\right) 
\nonumber \\
= & -e_a^\lambda e_{b \rho} \left(
(\partial_\mu\tilde{\omega}^{ab}_{\ \ \nu})-(\partial_\nu\tilde{\omega}^{ab}_{\ \ \mu})
-(\tilde{\omega}^{ac}_{\ \ \mu}\tilde{\omega}_{c\ \nu}^{\ b}
-\tilde{\omega}^{ac}_{\ \ \nu}\tilde{\omega}_{c\ \mu}^{\ b})
\right). \label{tildeR}
\end{align}
We also note some equations on traces of gamma matrices and their product with curvature and torsion tensors. 
\begin{align}
Tr(\gamma^{\mu\nu}) = 0, \ \ & Tr(\gamma^{\mu\nu}\gamma^{\rho\sigma}) = 4(g^{\mu\sigma}g^{\nu\rho}-g^{\mu\rho}g^{\nu\sigma}), \\
Tr\frac{1}{8}(\gamma^{\mu\nu}\gamma^{\rho\sigma})\tilde{R}_{\rho\sigma\mu\nu} 
= & \frac{4}{8}(g^{\mu\sigma}g^{\nu\rho}-g^{\mu\rho}g^{\nu\sigma})\tilde{R}_{\rho\sigma\mu\nu}
=-g^{\nu\sigma}\tilde{R}^\rho_{\ \sigma\rho\nu}= -\tilde{R}, \label{gamma2R}\\
Tr(-\frac{1}{16}\gamma^{\alpha\beta}\gamma^{\rho\sigma}T_{\mu\alpha\beta}T^\mu_{\ \rho\sigma}) = &
-\frac{4}{16}(g^{\alpha\sigma}g^{\beta\rho}-g^{\alpha\rho}g^{\beta\sigma})T_{\mu\alpha\beta}T^\mu_{\ \rho\sigma}
=\frac{1}{2}T^{\mu\rho\sigma}T_{\mu\rho\sigma}.
\end{align}
The Riemann curvature tensor $\tilde{R}^\lambda_{\ \rho\mu\nu}$, the Ricci tensor $\tilde{R}_{\rho\nu}$ and the 
curvature $\tilde{R}$ with torsion are related to those without torsion by
 \begin{align}
\tilde{R}^\lambda_{\ \rho\mu\nu} & = 
R^\lambda_{\ \rho\mu\nu}+\nabla_\mu Y^\lambda_{\ \rho\nu}-\nabla_\nu Y^\lambda_{\ \rho\mu}
+Y^\lambda_{\ \sigma\mu}Y^\sigma_{\ \rho\nu}-Y^\lambda_{\ \sigma\nu}Y^\sigma_{\ \rho\mu}, \label{tcurvet}\\
\tilde{R}_{\rho\nu} & =
\tilde{R}^\lambda_{\ \rho\lambda\nu} = R_{\rho\nu}+\nabla_\lambda Y^\lambda_{\ \rho\nu}-\nabla_\nu Y^\lambda_{\ \rho\lambda}
+Y^\lambda_{\ \sigma\lambda}Y^\sigma_{\ \rho\nu}-Y^\lambda_{\ \sigma\nu}Y^\sigma_{\ \rho\lambda}, \label{tRicci}\\
\tilde{R} & =R-2\nabla_\mu T^{\alpha\mu}_{\ \ \alpha}-T^{\alpha\mu}_{\ \ \ \alpha}T^\beta_{\ \mu\beta}
+\frac{1}{4}T^{\mu\nu\lambda}T_{\mu\nu\lambda}+\frac{1}{2}T^{\mu\nu\lambda}T_{\lambda\nu\mu} \nonumber \\
& = R + \nabla_\mu(\ovl{\psi}^\mu\gamma_\nu\psi^\nu)
-\frac{1}{4}\ovl{\psi}^\mu\gamma^\alpha\psi_\alpha\ovl{\psi}_\mu\gamma^\beta\psi_\beta
+\frac{1}{8}\ovl{\psi}^\nu\gamma^\mu\psi^\lambda\ovl{\psi}_\nu\gamma_\lambda\psi_\mu 
\nonumber \\
& + \frac{1}{16}\ovl{\psi}^\nu\gamma^\mu\psi^\lambda\ovl{\psi}_\nu\gamma_\mu\psi_\lambda. \label{tcurve}
\end{align} 
\renewcommand{\theequation}{B.\arabic{equation}}
\setcounter{equation}{0}
\section{Supertrace of $\mathbb{Z}$, $\mathbb{Z}\mathbb{Z}$, $\Omega_{\mu\nu}\Omega^{\mu\nu}$}
Using Eq.(\ref{Zphi}), (\ref{Zpsi}), the supertrace of the matrix $\mathbb{Z}$ of (\ref{matrixZ}) and 
$\mathbb{Z}\mathbb{Z}$ are given by 
\begin{align}
{\rm Str} \mathbb{Z} = & 4Z^{(\varphi)}-Tr Z^{(\psi)} = (1+4\lambda)\tilde{R} - \frac{1}{2}T^{\mu\rho\sigma}T_{\mu\rho\sigma}, \\
{\rm Str} \mathbb{Z}\mathbb{Z} = & 4 Z^{(\varphi)}Z^{(\varphi)}-Tr Z^{(\psi)}Z^{(\psi)} \nonumber \\
= & (4\lambda^2-\frac{1}{4})\tilde{R}^2-\tilde{R}_{\nu\rho}(-\tilde{R}^{\nu\rho}+\tilde{R}^{\rho\nu}) 
-\frac{1}{4}\tilde{R}_{\mu\nu\lambda\rho}(
 \tilde{R}^{\mu\nu\lambda\rho}+\tilde{R}^{\mu\lambda\rho\nu}+\tilde{R}^{\mu\rho\nu\lambda}
 ) \nonumber \\
& -\frac{1}{4}\tilde{R}_{\mu\nu\lambda\rho}(
 \tilde{R}^{\lambda\rho\mu\nu}+\tilde{R}^{\lambda\nu\rho\mu}+\tilde{R}^{\nu\rho\lambda\mu}
) 
-\frac{1}{2}(\tilde{\nabla}_\mu T^\mu_{\ \alpha\beta})(\tilde{\nabla}_\nu T^{\nu \beta\alpha}) \nonumber \\
& - \frac{1}{8}(
T^{\xi\mu\nu}T_{\xi\mu\nu}T_{\zeta\alpha\beta}T^{\zeta\alpha\beta}
+2T^{\xi\mu\nu}T_{\xi\alpha\beta}T_{\zeta\mu\nu}T^{\zeta\alpha\beta}
+4T^{\xi\mu\nu}T_{\xi\alpha\beta}T_{\zeta\mu}^{\ \ \alpha}T^{\zeta\beta}_{\ \ \nu}) \nonumber \\
& -(1+4\lambda)\tilde{R}\tilde{\nabla}_\alpha Q^\alpha
+2\tilde{R}_{\rho\nu} \tilde{\nabla}_\sigma T^{\sigma \nu\rho}
+(\frac{1}{2}+2\lambda)\tilde{R}Q_\mu Q^\mu 
+\frac{1}{4}\tilde{R}T_{\kappa\rho\sigma}T^{\kappa\rho\sigma} \nonumber \\
& -\tilde{R}_{\lambda\rho\mu\nu}T_\sigma^{\ \lambda\mu}T^{\sigma\rho\nu} 
 +\frac{1}{2}\tilde{R}_{\lambda\rho\mu\nu}T_\sigma^{\mu\nu}T^{\sigma\lambda\rho} 
 -\frac{1}{2}(\tilde{\nabla}_\mu Q^\mu)T_{\nu\rho\sigma}T^{\nu\rho\sigma} \nonumber \\
& -\frac{1}{4}Q^\alpha Q_\alpha T_{\kappa\mu\nu}T^{\kappa\mu\nu}. \label{StrZZ}
\end{align}
In the same way, using (\ref{Omegaphi}), (\ref{Omegapsi}), we obtain the supertrace of $\Omega_{\mu\nu}\Omega^{\mu\nu}$ as 
follows:
\begin{align}
\lefteqn{{\rm Str}\Omega_{\mu\nu}\Omega^{\mu\nu}} \nonumber \\
= &
4\Omega^{(\varphi)}_{\mu\nu}\Omega_{(\varphi)}^{\mu\nu}-Tr \Omega_{\mu\nu}^{(\psi)}\Omega^{\mu\nu}_{(\psi)}
\nonumber \\
= & \frac{1}{2}\tilde{R}_{\mu\nu\alpha\beta}\tilde{R}^{\mu\nu\alpha\beta}
+g^{\mu\nu}g^{\rho\sigma}\left(
(\tilde{\nabla}_\mu T_\rho^{\ \alpha\beta})(\tilde{\nabla}_\nu T_{\sigma\alpha\beta})
-(\tilde{\nabla}_\mu T_{\rho\alpha\beta})(\tilde{\nabla}_\sigma T_\nu^{\ \alpha\beta})
\right)\nonumber \\
& +\frac{1}{4}(T_\alpha^{\ \mu\nu}T^{\alpha\lambda}_{\ \ \ \nu}T_{\beta\mu}^{\ \ \ \sigma}T^\beta_{\ \lambda\sigma}
-T_\alpha^{\ \mu\nu}T^{\alpha\lambda}_{\ \ \ \sigma}T_{\beta\mu}^{\ \ \ \sigma}T^\beta_{\ \lambda\nu}
-2T_\alpha^{\ \mu\nu}T^{\alpha\lambda}_{\ \ \ \sigma}T^\beta_{\ \mu\nu}T_{\beta\lambda}^{\ \ \ \sigma}
\nonumber \\
& -32T_\alpha^{\ \mu\nu}T^\alpha_{\ \beta\rho}T_\mu^{\ \lambda\beta}T_{\nu\lambda}^{\ \ \ \rho}
) 
+4\tilde{R}^{\lambda\rho\mu\nu}\tilde{\nabla}_\mu T_{\nu\lambda\rho}
\nonumber \\
& -8\tilde{R}_{\alpha\beta\mu\nu}T^{\mu\sigma\alpha}T_{\ \sigma}^{\nu\ \beta}
-\tilde{R}_{\alpha\beta\mu\nu}T^{\sigma\mu\nu}T_\sigma^{\ \alpha\beta}
-16(\tilde{\nabla}_\mu T_\nu^{\ \alpha\beta})T^{\mu\sigma}_{\ \ \alpha}T^\nu_{\ \ \sigma\beta}\nonumber \\
& -2(\tilde{\nabla}_\mu T_\nu^{\ \alpha\beta})T^{\lambda\mu\nu}T_{\lambda\alpha\beta}. \label{StrOmegaOmega}
\end{align}
When we develop terms of the second power of the Ricci and Riemann curvature tensors in the equation (\ref{StrZZ}), 
we can use equations as follows:
\begin{align}
\lefteqn{-\frac{1}{2}\tilde{R}_{\nu\rho}(-\tilde{R}^{\nu\rho}+\tilde{R}^{\rho\nu})+\tilde{R}_{\mu\nu}\tilde{\nabla}_\lambda
T^{\lambda\nu\mu}-\frac{1}{4}(\tilde{\nabla}_\mu T^\mu_{\ \alpha\beta})(\tilde{\nabla}_\nu T^{\nu\beta\alpha}) }\nonumber \\
& =  \frac{1}{4}(\tilde{\nabla}_\mu Q_\nu - \tilde{\nabla}_\nu Q_\mu-Q_\alpha T^\alpha_{\ \mu\nu})
(\tilde{\nabla}^\mu Q^\nu - \tilde{\nabla}^\nu Q^\mu-Q_\beta T^{\beta\mu\nu}) \nonumber \\
& = \frac{1}{2}\left((\tilde{\nabla}_\mu Q_\nu)(\tilde{\nabla}^\mu Q^\nu)-(\tilde{\nabla}_\mu Q_\nu)(\tilde{\nabla}^\nu Q^\mu) -(\tilde{\nabla}_\mu Q_\nu)Q_\alpha T^{\alpha\mu\nu}\right) 
+Q_\alpha T^{\alpha\mu\nu}Q_\beta T^\beta_{\ \mu\nu}.
\end{align}
\begin{align}
\lefteqn{-\frac{1}{8}\tilde{R}_{\mu\nu\lambda\rho}(
\tilde{R}^{\mu\nu\lambda\rho}+\tilde{R}^{\mu\lambda\rho\nu}+\tilde{R}^{\mu\rho\nu\lambda}
 +\tilde{R}^{\lambda\rho\mu\nu}+\tilde{R}^{\lambda\nu\rho\mu}+\tilde{R}^{\nu\rho\lambda\mu}
)}\nonumber \\
= & -\frac{1}{16}(\tilde{\nabla}_\mu T_{\rho\alpha\beta})(\tilde{\nabla}^\mu T^{\rho\alpha\beta})
-\frac{1}{8}(\tilde{\nabla}_\mu T_{\rho\alpha\beta})(\tilde{\nabla}^\alpha T^{\rho\beta\mu})
+\frac{1}{16}(\tilde{\nabla}_\mu T_{\rho\alpha\beta})(\tilde{\nabla}^\rho T^{\mu\alpha\beta}) \nonumber \\
& -\frac{1}{8}(\tilde{\nabla}_\mu T_{\rho\alpha\beta})(\tilde{\nabla}^\alpha T^{\beta\mu\rho})
-\frac{1}{8}(\tilde{\nabla}_\mu T_{\rho\alpha\beta})(\tilde{\nabla}^\mu T^{\beta\rho\alpha})
-\frac{1}{8}(\tilde{\nabla}_\mu T_{\rho\alpha\beta})(\tilde{\nabla}^\rho T^{\beta\alpha\mu})
\nonumber \\
& +\frac{1}{8}(\tilde{\nabla}_\mu T_{\rho\alpha\beta})(\tilde{\nabla}^\alpha T^{\mu\beta\rho}) 
-\frac{1}{8}(\tilde{\nabla}_\mu T_\nu^{\ \alpha\beta})(
  2T^{\sigma\mu}_{\ \ \ \alpha}T_{\beta\sigma}^{\ \ \ \nu}+ 2T^{\sigma\nu}_{\ \ \ \alpha}T_{\beta\ \sigma}^{\ \mu}
  +2T^{\sigma\nu\mu}T_{\alpha\beta\sigma}
\nonumber \\
& -T^{\mu\nu\sigma}T_{\sigma\alpha\beta}
  -2T^{\sigma\nu}_{\ \ \ \alpha}T^\mu_{\ \beta\sigma}+2T^{\sigma\mu}_{\ \ \ \alpha}T^\nu_{\ \beta\sigma})
+\frac{3}{8}(\tilde{\nabla}_\mu T_\nu^{\ \alpha\beta})T_{\sigma\alpha\beta}T^{\nu\mu\sigma} \nonumber \\
& +\frac{1}{8}
T^{\xi\mu\nu}T_{\alpha\xi\beta}(T^{\zeta\ \beta}_{\ \nu}T_{\mu\zeta}^{\ \ \ \alpha}
+T^{\zeta\alpha\beta}T_{\mu\nu\zeta}
+T^{\zeta\alpha}_{\ \ \ \nu}T_{\mu\zeta}^{\ \ \ \beta} 
-T^{\zeta\ \alpha}_{\ \mu}T_{\ \zeta\nu}^{\beta}
-\frac{1}{2}T^\zeta_{\ \nu\mu}T_{\beta\zeta\alpha} \nonumber \\
& -T^{\zeta\ \beta}_{\ \mu}T_{\ \zeta\nu}^{\alpha}
-\frac{1}{2}T^\zeta_{\ \nu\mu}T^{\alpha\ \beta}_{\ \zeta} 
-T_{\zeta\mu\nu}T^{\alpha\zeta\beta}-2T^{\zeta\beta\nu}T^\alpha_{\ \mu\zeta}).
\end{align}
\end{document}